  \documentclass[preprint2]{aastex}

\def\gsim{ \lower .75ex \hbox{$\sim$} \llap{\raise .27ex \hbox{$>$}} }
\def\lsim{ \lower .75ex\hbox{$\sim$} \llap{\raise .27ex \hbox{$<$}} }

\begin{document}

\title{The Luminosity and Angular Distributions of Long-Duration GRBs}

\author{Dafne Guetta\altaffilmark{1,}\altaffilmark{2},
Tsvi Piran\altaffilmark{1}
and Eli Waxman\altaffilmark{3}}

\altaffiltext{1}{Racah Institute for Physics, The Hebrew
University, Jerusalem 91904, Israel} \altaffiltext{2}{JILA,
University of Colorado, Boulder, CO 80309, USA}
\altaffiltext{3}{Department of Condensed Matter Physics, Weizmann
Institute, Rehovot 76100, Israel}

\begin{abstract}

The realization that the total energy of GRBs is correlated with
their jet break angles motivates the search for a similar relation
between the peak luminosity, $L$, and the jet break angles,
$L\propto\theta^{-2}$. Such a relation implies that the GRB
luminosity function determines the angular distribution. We
re-derive the GRB luminosity function using the BATSE peak flux
distribution and compare the predicted distribution with the
observed redshift distribution. The luminosity function can be
approximated by a broken power law with a break peak luminosity
of $4.4 \times 10^{51}$ erg/sec, a typical jet angle of 0.12 rad
and a local GRB rate of $0.44\,h_{65}^{3}$~Gpc$^{-3}$yr$^{-1}$.
The angular distribution implied by $L\propto\theta^{-2}$ agrees
well with the observed one, and implies a correction factor to the
local rate due to beaming of $75\pm25$ (instead of 500 as commonly
used).  The inferred overall local GRB rate is
$33\pm11\,h_{65}^{3}$~Gpc$^{-3}$yr$^{-1}$.  The luminosity
function and angle distribution obtained within the universal
structured jet model, where the angular distribution is
essentially $\propto \theta$ and hence the luminosity function
must be $\propto L^{-2}$, deviate from the observations at low
peak fluxes and, correspondingly, at large angles. The
corresponding correction factor for the universal structure jet is
$\sim 20 \pm 10$.

\end{abstract}
\keywords{cosmology:observations-gamma rays:bursts}

\section{Introduction}

In spite of great progress, one of the missing links in our
understanding of GRBs is a knowledge of the luminosity function
and the rate of GRBs. The realization that GRB are beamed raised
the question of what is the angular distribution. The discovery of
Energy - angle relation (Frail et al., 2001; Panaitescu \& Kumar
2001) suggested that the luminosity function and the angular
distributions should be also related.

The number of GRBs with an observed afterglow is still rather
limited. Redshift has been measured only for a fraction of these
bursts and the selection effects that arise in this sample are not
clear. Therefore, at present we cannot derive directly the  GRB
luminosity function. A possible method to derive the luminosity
function makes use of a luminosity indicator. Several such
indicators have been suggested (Norris, Marani \& Bonnell 2000;
Fenimore \& Ramirez-Ruiz 2000) but the robustness of those
indicators is not yet clear.  In a recent paper  Schmidt (2003)
shows that a luminosity function obtained using luminosity
indicators fits the observations very poorly. An alternative is to
consider the observed peak flux distribution and fit possible
luminosity functions and burst rate distributions (Piran 1992,
Cohen \& Piran 1995, Fenimore \& Bloom 1995, Loredo \& Wasserman
1995, Horack \& Hakkila 1997, Loredo \& Wasserman 1998, Piran
1999, Schmidt 1999, Schmidt 2001, Sethi \& Bhargavi 2001) This can
be done by obtaining a best fit for the whole distribution or
simply by just fitting the lowest moment, namely, $\langle
V/V_{max} \rangle$.

Evidence of jetted GRBs arises  from long term radio observations
(Waxman, Kulkarni \& Frail 1998) and from observations of
achromatic breaks in the afterglow light curves (Rhoads 1997,
Sari, Piran \& Halpern 1999). However the structure of these jets
is still an open question. The two leading models are (1) the
uniform jet model, where the energy per solid angle, $\epsilon$,
is roughly constant within some finite opening angle, $\theta$,
and sharply drops outside of $\theta$, and (2) the universal
structured jet (USJ) model (Rossi, Lazzati \& Rees 2002), where
all GRB jets are intrinsically identical, and $\epsilon$ drops as
the inverse square of angle from the jet axis.
 Within the uniform jet model, the observed break
corresponds to the jet opening angle. Frail et al. (2001) and
Panaitescu and Kumar (2001) have estimated the opening angles
$\theta$ for several GRBs  with known redshifts. They find that
the total gamma-ray energy release, when corrected for beaming as
inferred from the afterglow light curves, is clustered. A recent
analysis (Bloom, Frail \& Kulkarni, 2003) on a larger sample
confirmed this clustering around $\sim 1.3\times 10^{51}$ergs.
Within the USJ model the jet break corresponds to the viewing
angle and the energy-angle relation given above implies  that the
luminosity distribution within the jet is proportional to
$\theta^{-2}$  (Rossi, Lazzati \& Rees 2002) and therefore the
luminosity function has the form $\Phi(L)\propto L^{-2}$  (Perna,
Sari \& Frail 2003).
%that the
%energy distribution within the jet is proportional to
%$\theta^{-2}$ (Rossi et al. 2002).
Frail et al. (2001) also derive
the observed $\theta$ distribution. Taking into account the fact
that for every observed burst there are $f_b^{-1} \equiv
(\theta^2/2)^{-1}$ that are not observed they derived the true
$\theta$ distribution and used it to estimate the average
``beaming factor'',  $\langle f_b^{-1} \rangle \simeq 500$. Then
they multiply this factor times the local (isotropic estimated)
GRB rate derived by Schmidt (2001) and obtain the true local rate
of GRB.

We aim here at obtaining the combined L-$\theta$ distribution
function, $\Phi(L,\theta)$  in the uniform jet model. The present
data is not sufficient to accurately constrain $\Phi(L,\theta)$.
We therefore must make some assumptions.  First we estimate the
{\it isotropic} luminosity function, $\Phi(L)=\int d\theta
f_b(\theta)\Phi(L,\theta)$, by fitting the peak flux distribution
using a simple parametrization of this function. When considering
the angular distribution we also examine (following Perna, Sari \&
Frail 2003) the USJ model  and show that the implied peak flux
distribution is somewhat inconsistent with the observed one, with
the problem most severe at the low end of the peak luminosity
distribution. 

We find that the data support $L\times
\theta^2\sim {\rm const}$ (see also Van Putten \& Regimbau 2003).
There is a clustering of the true peak luminosity around $\sim 3.2
\times 10^{49}$ erg/s, but note that the distribution is not a
narrow delta-function. We notice then that once we have such a
relationship the luminosity function implies an angular
distribution function.
This is true even when some spread in the value of
$L\times \theta^2$ is taken into account, reflecting the spread observed
in current data.
The resulting angular and redshift
distributions are consistent with the observations. Our analysis
is well motivated but approximate. Due to width of the luminosity
angle distribution and given the data quality and selection
effects, we believe there is no point in rigorous maximal
likelihood analysis and the like. We give approximate estimates of
the model parameters and approximate uncertainties.

The paper is structured in two main parts. In the first part we
rederive (following Schmidt, 1999) the GRB luminosity function. In
the second part we consider the angular distribution that follows
from the luminosity function and the peak luminosity - angle
relation. We compare the predicted angular distribution with the
observed one. Using this distribution we re-calculate the
correction to the rate of GRBs due to beaming. Unlike Frail et
al., (2001) we use a weighted average of the angular distribution.
This yields a significantly smaller correction factor. We also
estimate a the correction factor needed for the USJ model. Note
that this revised correction factor does not depend on the
relation between the peak luminosity and the angle.

The observed sample we use to explore the peak luminosity - angle
relation is small and its  selection effects  are hard to
quantify. Therefore it is difficult to give robust conclusions
from our analysis at this stage.  However, future missions and in
particular SWIFT  will allow us, hopefully in the near future,  to
use this procedure and test our conclusions with larger and less
biased samples.

\section{Luminosity function  from the BATSE sample}

We consider   all the long GRBs ($T_{90}> 2$sec) (Kouveliotou et
al., 1993), detected while the BATSE onboard trigger  (Paciesas et
al. 1999) was set for 5.5 $\sigma$ over background in at least two
detectors, in the energy range 50-300 keV. Among those we took the
bursts for which $C_{\rm max}/C_{\rm min} \geq 1$ at the 1024 ms
timescale, where
 $C_{\rm max}$ is the count rate in the second brightest illuminated detector
and $C_{\rm min}$ is the minimum detectable rate.  Using this
sample of 595 GRBs we find $\langle V/V_{\rm max} \rangle=0.294$.

The method used to derive the luminosity function is essentially
the same of Schmidt (1999). We consider a broken power law with
lower and  upper limits, $1/\Delta_1$ and $\Delta_2$, respectively.
The local luminosity function of GRB peak luminosities $L$,
defined as the co-moving space density of GRBs in the interval
$\log L$ to $\log L + d\log L$ is:
\begin{equation}
\label{Lfun}
\Phi_o(L)=c_o
\left\{ \begin{array}{ll}
(L/L^*)^{\alpha} &  L^*/\Delta_1 < L < L^* \\
(L/L^*)^{\beta} & L^* < L < \Delta_2 L^*
\end{array}
\right. \;,
\end{equation}
where $c_o$ is a normalization constant so that the integral over
the luminosity function equals unity. We stress that this
luminosity function is the ``isotropic-equivalent" luminosity function. I.e.
it does not include a correction factor due to the fact that GRBs
are beamed.

Following Schmidt (2001) we employ the parametrization of Porciani
\& Madau (2001), in particular, their SFR model SF2.
\begin{eqnarray}
\nonumber
\label{SFR} R_{GRB}(z) = R_{\rm SF2}(z) \\
=  \rho_0 \frac{23
\exp(3.4z)}{\exp(3.4z)+22}
\, F(z,\Omega_M,\Omega_{\Lambda})
\end{eqnarray}
where $\rho_0$ is the GRB rate at $z=0$ and
$F(z,\Omega_M,\Omega_{\Lambda})=
[\Omega_M(1+z)^3+\Omega_k(1+z)^2+\Omega_{\Lambda}]^{1/2}/(1+z)^{3/2}$.

We consider also the
Rowan-Robinson SFR (Rowan-Robinson 1999: RR-SFR) that can be
fitted with the expression
\begin{equation}
\label{RR}
R_{GRB}(z) = \rho_0
\left\{ \begin{array}{ll}
10^{0.75 z} & z<1 \nonumber \\
10^{0.75 z_{\rm peak}} & z>1.
\end{array}
\right.
\end{equation}
For given values of the parameters
$\Delta_1,\,\Delta_2,\alpha,\beta$ we  determine $L^*$ so that the
predicted value $\langle V/V_{\rm max} \rangle$ equals the
observed one
and we find the local rate of GRB, $\rho_0$, from the observed
number of GRBs. We then consider the range of parameters for which
there is a reasonable fit to the peak flux distribution (see
Figure 1). We use the cosmological parameters $H_0 = 65~$ km
s$^{-1}$ Mpc$^{-1}$, $\Omega_M = 0.3$, and $\Omega_{\Lambda} =
0.7$.
% and we consider a spectral index $-1.6$ as representing the
%spectrum for all bursts.

The modeling procedure involves the derivation of the peak flux
P(L,z) of a GRB of peak luminosity L observed at redshift z:
\begin{equation}
\label{peak}
P(L,z)=\frac{L}{4\pi D_L^2(z)}
\frac{C(E_1(1+z),E_2(1+z))}{C(E_1,E_2)}
\end{equation}
where $ D_L(z)$ is the bolometric luminosity distance and
$C(E_1,E_2)$ is the integral of the spectral energy distribution
between $E_1 = 50$ keV and $E_2=300$ keV. Schmidt (2001) finds
that the median value of the spectral photon index in the
50-300keV band for the long bursts sample is -1.6 and this can be
used for a simplified k-correction. We use this value for our
analysis.

Objects with luminosity $L$ observed by BATSE with a flux limit
$P_{\rm lim}$ are detectable to a maximum redshift $z_{\rm
max}(L,P_{\rm lim})$  that can be derived from Eq. \ref{peak}. The
limiting flux has a distribution $G(P_{\rm lim}$) that can be
obtained from the distribution of $C_{\rm min}$ of the BATSE
catalog. Considering four main representative intervals we get
that 6\% of the sample has $P_{\rm lim}\sim 0.20 $  ph
cm$^{-2}$s$^{-1}$, 18\% has $P_{\rm lim}\sim 0.25 $  ph
cm$^{-2}$s$^{-1}$, 52\% has $P_{\rm lim}\sim 0.27 $  ph
cm$^{-2}$s$^{-1}$ and 24\% has $,P_{\rm lim}\sim 0.32 $  ph
cm$^{-2}$s$^{-1}$.  The inclusion of this variation of $P_{\rm
lim}$, and the implied different samples, is the main difference
between our analysis of the luminosity function and Schmidt's.

The number of bursts with a peak flux $>P$ is given by:
%\begin{equation}
%N(>P)=\int\Phi_o(L)d\log L \int G(P_{\rm lim}) dP_{\rm lim}
%\int_0^{z_{max}(L,P/P_{\rm lim})} \frac{R_{GRB}(z)}{1+z}
%\frac{dV(z)}{dz}dz \
%\end{equation}
\begin{eqnarray}
\nonumber
  N(>P)=\int\Phi_o(L)d\log L \int G(P_{\rm lim}) dP_{\rm lim} \\
   \int_0^{z_{max}(L,P/P_{\rm lim})} \frac{R_{GRB}(z)}{1+z}
\frac{dV(z)}{dz}dz \
\end{eqnarray}
where the factor  $(1+z)^{-1}$ accounts for the cosmological time
dilation and $dV(z)/dz$ is the comoving volume element. We
determine $L^*$ so that $\langle V/V_{\rm max} \rangle=0.294$.

If we approximate Schmidt's results (Schmidt 2001) the luminosity
function can be characterized as two power laws of slopes
$\alpha=-0.6$ and $\beta=-2$, with $\Delta_1=30$ and $\Delta_2=10$
and with an isotropic-equivalent break peak luminosity of $L^*\sim
3.2\times10^{51}$ erg/s and a local GRB  $\rho_0 \sim 0.5$
Gpc$^{-3}$yr$^{-1}$. Using these values we show in Figure 1  that
the predicted logN-log$(P/P_{\rm lim})$ distribution doesn't agree
with the observed logN-log$(C_{\rm max}/C_{\rm min})$. In
principle one should perform a maximum likelihood analysis to
obtain the best values of the luminosity parameters (and even the
GRB rate parameters). However, we feel that a simpler approach is
sufficient for our purpose, especially in view of the small size
of the sample used in the later part of the analysis. We simply
vary $\alpha$ and $\beta$, keeping $\Delta_1=30$ and $\Delta_2=10$
and inspect the quality of the fit to the observed
logN-log$(C_{\rm max}/C_{\rm min})$ distribution. To obtain the
local rate of GRBs per unit volume, $\rho_0$ we need to  estimate
the effective full-sky coverage of our GRB sample. We find 595
events in this sample that were detected over 1386 days in the
50-300 keV channel of BATSE, with a sky exposure of 48\% . We also
take into account that this sample is 47\% of the long GRBs.

In Figure 1 we show a comparison of the observed logN-log$(C_{\rm
max}/C_{\rm min})$ with several predicted logN-log$(P/P_{\rm
lim})$ distributions obtained with the RR-SFR  and with a non
evolving luminosity function of the form given in Eq. \ref{Lfun}
(the distributions are similar for the SF2). We find reasonable
fits with $-0.6<\alpha<-0.1$ and $-3<\beta<-2$ taking into account
that $\alpha$ and $\beta$ are correlated, namely, $\alpha$ from
-0.1 to -0.6 require decreasing $\beta$ from -2 to -3. As we can
see from Table 1, among the different acceptable models $L^*$
varies by a factor of $\sim 3$.  $\rho_0$ varies by a factor of
$\sim 2$ within a given SFR and as expected it is larger by a
factor of $\sim 2$ in the RR model than in SF2.

Since random errors in a cumulative distribution like $N(>P)$
propagate in an unknown way, we present in Figure 2 our results
for the differential distributions $n(P)\equiv dN/dP$. We find
reasonable fits for the range of $\alpha$ and $\beta$ given above.

For our analysis we used 595 long GRBs detected at a time scale
of 1024 ms from the BATSE catalog. A much larger sample is now
available in the GUSBAD catalog which lists 2204 GRBs at a time
scale of 1024 ms \\
(http://www.astro.caltech.edu/~mxs/grb/GUSBAD). We find that our
main results remain the same if we consider this catalog.

\begin{table*}[t]
\begin{center}
\begin{tabular}{lllll}
\hline SFR & $\alpha$ & $\beta$ & $L^*$ (erg/s)
& $ \rho_0$ (Gpc$^{-3}$yr$^{-1}$)\\
\hline
SF2 & -0.1 & -2 & $6.3 \times 10^{51}$ & 0.18 \\
SF2 & -0.6 & -3 & $1.6 \times 10^{52}$& 0.16 \\
RR-SFR & -0.1 & -2 & $4.4 \times 10^{51}$& 0.44 \\
RR-SFR & -0.6 & -3 & $1.1\times  10^{52}$ & 0.19 \\
\hline
\end{tabular}
\caption{Limiting values of $\alpha$ and $\beta$ ranges needed to
reproduce the observed $(C_{\rm max}/C_{\rm min})$ distribution
and the implied values of $L^*$ and $\rho_0$ for the two SFRs
considered in the paper.}
\end{center}
\label{tab:fits}
\end{table*}

\begin{figure}[a]
%\centering \noindent
{\par\centering \resizebox*{0.85\columnwidth}{!}{\includegraphics
%[width=8cm,height=8cm]
{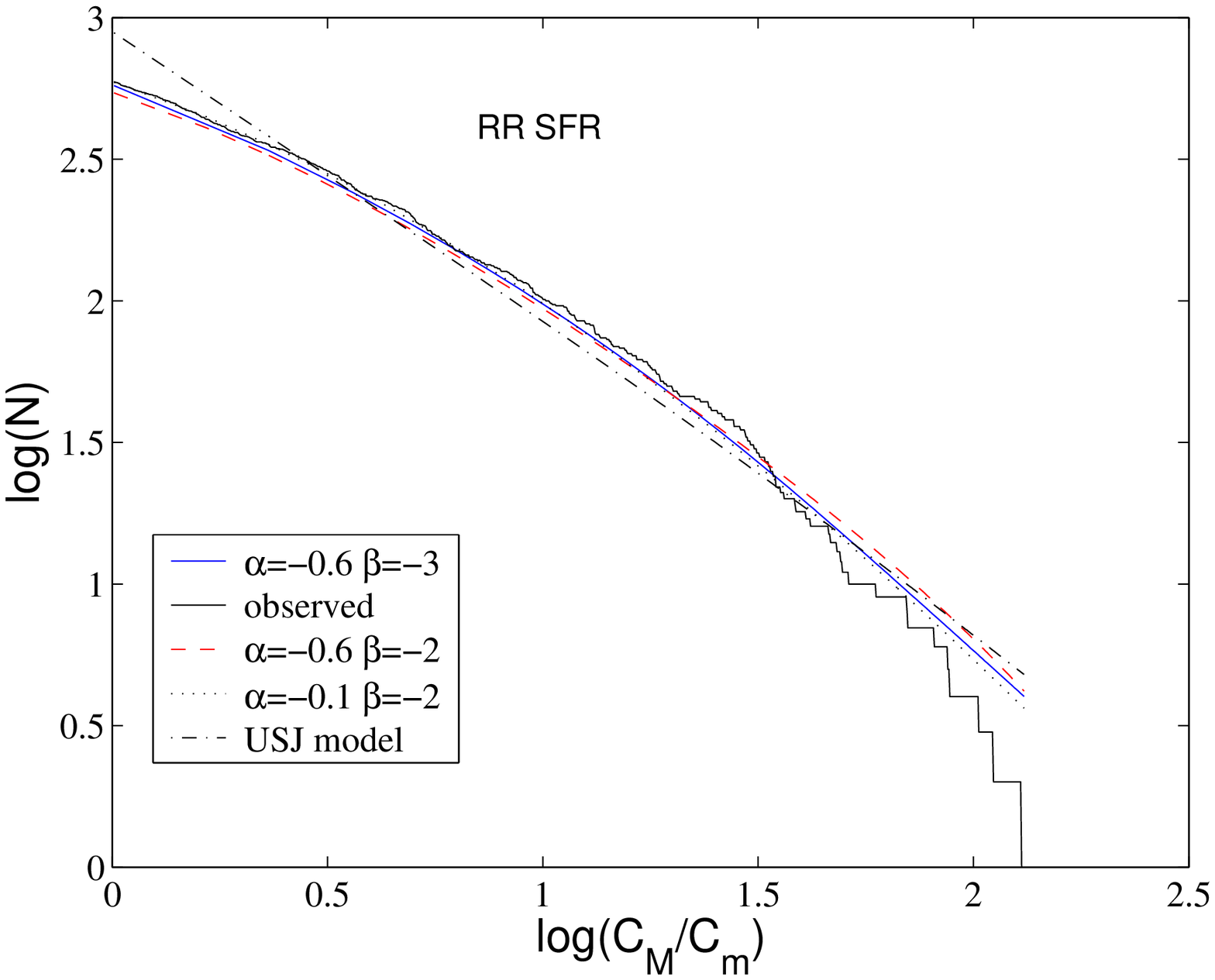}} \par}
 \caption{\label{fig1} The predicted
logN-log(P/P$_{\rm lim}$) distribution for different values of
$\alpha$ and $\beta$ with a RR-SFR vs. the  observed
logN-log(C$_{\rm max}$/C$_{\rm min}$) taken from the BATSE
catalog. We also  plot the predicted logN-log(P/P$_{\rm lim}$)
distribution in the  USJ jet model, for which the inconsistency at
the low peak flux range is apparent.}
\end{figure}

\begin{figure}[b]
%\centering \noindent
{\par\centering \resizebox*{0.85\columnwidth}{!}{\includegraphics
%[width=8cm,height=8cm]
{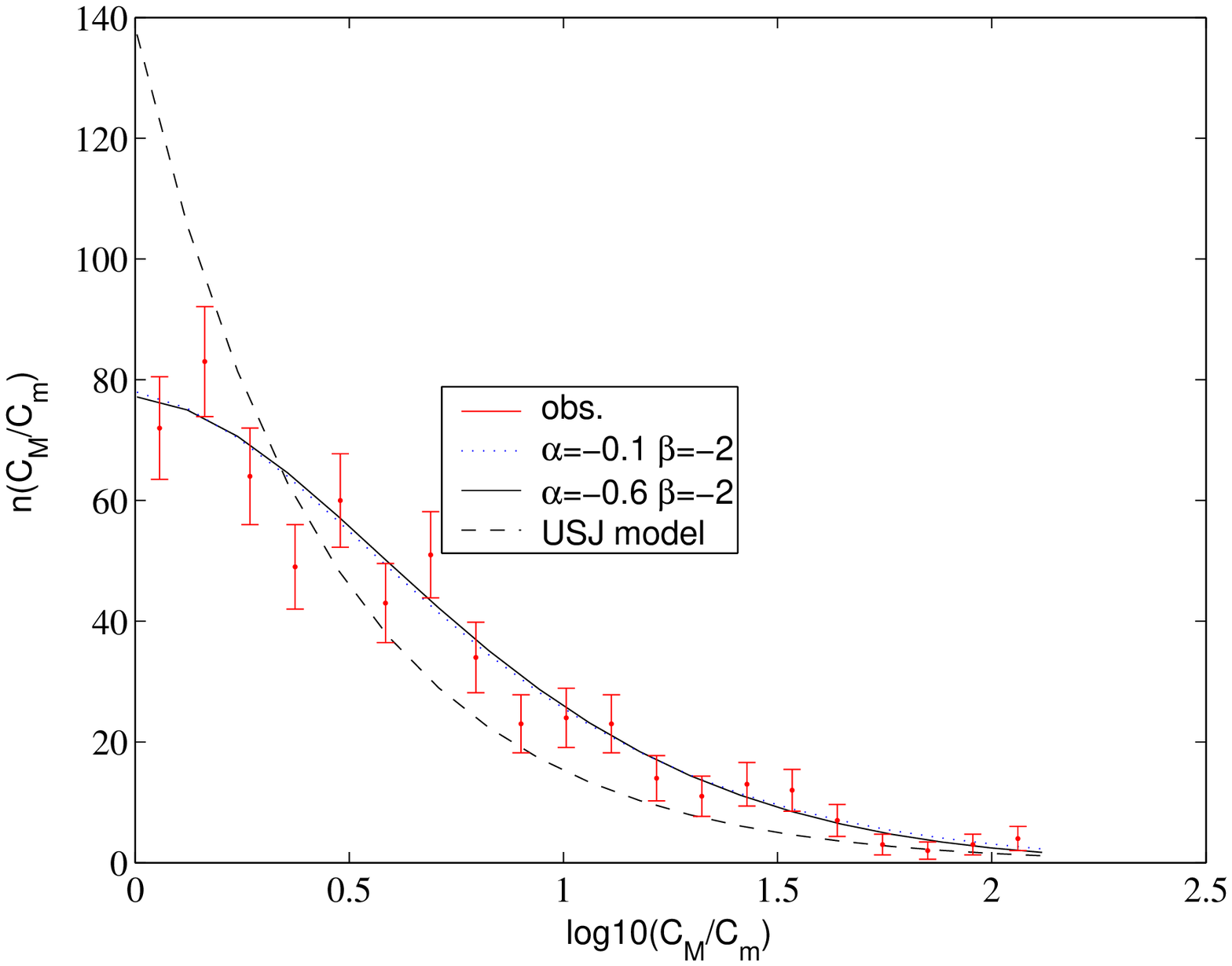}} \par}
 \caption{\label{fig1} The predicted
differential distribution, $n(P/P{\rm lim})$, for different values
of $\alpha$ and $\beta$ with a RR-SFR vs. the  observed n(C$_{\rm
max}$/C$_{\rm min}$) taken from the BATSE catalog. We also  plot
the predicted differential distribution in the  USJ jet model, for
which the inconsistency at the low peak flux range is apparent.}
\end{figure}

From  Figures 1 and 2 it is clear that the best set of values that
reproduce the logN-log$(C_{\rm max}/C_{\rm min})$ distribution is
$\alpha=-0.1$ and $\beta=-2$ and these will be the values used in
the rest of our paper. Our results are  slightly different from
Schmidt's
 mainly because we consider a different sample of GRBs
and the quantity $P/P_{\rm lim}$ (which correspond to the observed
$(C_{\rm max}/C_{\rm min})$) instead of the flux $P$ used by
Schmidt (2001). In this way we take into account the fact that
different bursts are detected with different flux thresholds.

So far we have used a parametric fit for the luminosity function.
Within the  USJ jet model, with $L \propto \theta^{-2}$ as forced
by the energy-angle relation, we have a simple luminosity
function: $\Phi_{sj} \propto L^{-2}$ (Perna et al. 2003).  In
Figures 1 and 2  we also compare the predicted logN-log$(P/P_{\rm
lim})$ distribution that arises from this luminosity function with
the RR-SFR with the observed logN-log$(C_{\rm max}/C_{\rm min})$
distribution. We find that within the USJ scenario the
distribution is inconsistent with the data. The problem arises
mostly for low peak fluxes. It is important to note, however, that our
method of GRB selection does not introduce a bias against the inclusion of
such low peak flux bursts. A ``quasi-universal" jet (Zhang et al. 2003)
obviously allows more freedom in fitting the data and may be
consistent.

We can use now the luminosity function to derive the observed
redshift distribution:
\begin{equation}
\label{redshift} N(z)= \frac{R_{GRB}(z)}{1+z} \frac{dV(z)}{dz}
\int_{L_{\rm min}(P_{\rm lim},z)}^{L_{\rm max}} \Phi_o(L)d\log L \
 ,
\end{equation}
where $L_{\rm min}(P_{\rm lim})$ is the luminosity corresponding
to the minimum peak flux $P_{\rm lim}$ for a burst at redshift z
and $L_{\rm max}=L^*\times \Delta_2=10\,L^*$. This  minimal peak
flux corresponds to the sensitivity of the gamma-ray burst
detector used.  We use,  $P_{\rm lim}\sim 0.5-1 $ ph cm$^{-2}
s^{-1}$, which is  roughly the limiting flux for the GRBM on
BeppoSAX (Guidorzi PhD thesis). We compare this distribution with
the observed distribution of all the bursts with an available
redshift: 32 bursts  from http://www.mpe.mpg.de/~jcg/grbgen.html
(excluding GRB980425 with $z=0.0085$). It is hard to quantify the
selection effects that arise in the determination of the redshift.
The problem is most severe in the range $1.3<z<2.5$ where  visual
spectroscopy (Hogg \& Fruchter, 1999). Following Hogg \& Fruchter
(1999) we consider all the GRB with optical afterglow but without
a measured redshift to be in this redshift range $1.3<z<2.5$.
Using all the bursts in http://www.mpe.mpg.de/~jcg/grbgen.html we
have a sample of 46 GRBs.

A comparison of  the predicted cumulative redshift distribution
with the observed one, (see Figure 3), reveals that with SF2 we
obtain  fewer bursts at low redshift than the observed ones. This
happens even after taking into account the selection effects. In
fact a Kolmogorov-Smirnoff test (KS) shows that the two
distributions are only marginally compatible (5\% level). The
RR-SFR seems to reproduce better the redshift distribution, as the
KS test shows that they are comparable at the level of 20\%.
Therefore  we will focus on this distribution in the rest of the
analysis.

\begin{figure}
%\centering \noindent
{\par\centering \resizebox*{0.95\columnwidth}{!}{\includegraphics
%[width=8cm,height=8cm]
{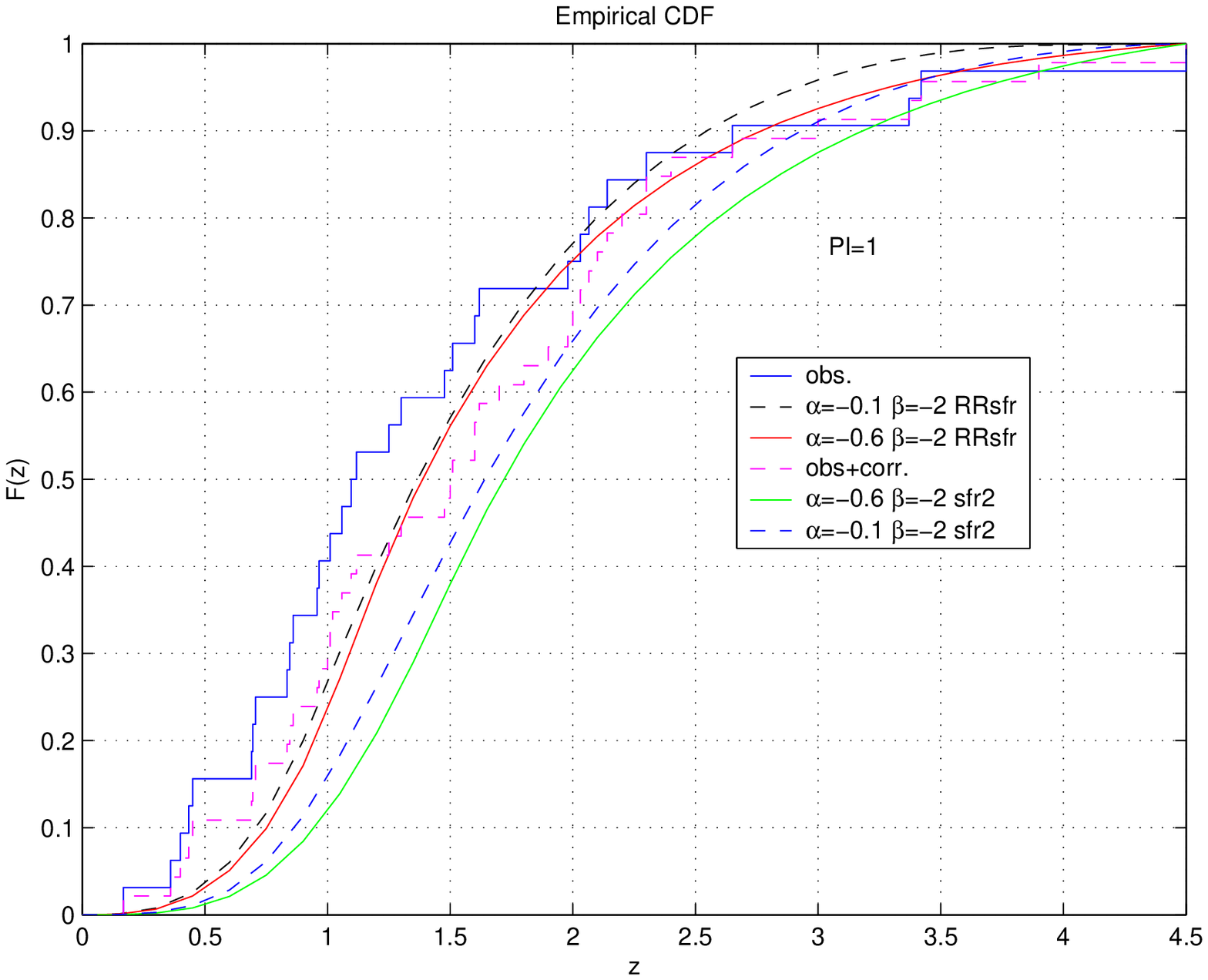}} \par} \caption{\label{fig2} The predicted
cumulative distributions of the GRBs redshift for the two SFR for
$\alpha=-0.6$ (solid line) and our $\alpha=-0.1$ (dashed line).
The histograms show the observed cumulative redshift distribution
and a distribution where selection effects are taken into account
assuming that all the GRB with no redshift but with optical
afterglow lie in the range $1.3<z<2.5$. }
\end{figure}

\section{Distribution of opening angles}

We turn now to the peak luminosity - angle relation, which is
analogous to the fluence - angle relation (Frail et al., 2001;
Panaitescu \& Kumar 2001).  We consider a sample of 9 GRBs of the
BATSE 4B catalog for which it has been possible to determine the
redshift and the jet opening angle, $\theta$ (Bloom et al. 2003).
In order to estimate the peak luminosity we need the peak fluxes.
The advantage in taking only the BATSE GRBs is that we have all
the peak fluxes estimated in the same way (averaged over the 1.024
sec BATSE trigger in the energy range 50-300 keV)

For all the bursts of this sample we estimate the
peak luminosity along the jet of opening angle $\theta$ as:
\begin{equation}
\label{ljet}
L_j=f_b L.
\end{equation}

In our analysis we consider the angles up to the maximum value
implied by the observations ($\sim 0.7$ rad). Within this range of
$\theta$ the beaming factor can be approximated as
$f_b=(1-\cos\theta)\sim\theta^2/2$. In the upper panel of Figure 4
(left side) we show the distribution of $L_j$ obtained from this
sample. Comparing this distribution with the isotropic one shown
in the lower panel of the same figure (left side) we note that the
peak luminosity, when corrected for beaming, is clustered around a
value $L_j\sim 10^{49.5}$ erg/s. A similar result, shown in Figure
4 (right side), is obtained when we consider a larger sample, of
19 bursts with an angular estimate from 0.05 to 0.7 radians. Here
there are larger uncertainties in the determination of the peak
luminosity, as the bursts were detected by different detectors
with different averaging time and different spectral resolution.
We have extrapolated the GRB fluxes to the BATSE range 50-300 keV
using the method described in Sethi \& Bhargavi (2001)
reelaborated for our spectrum.  The redshifts and fluxes were
taken from the table given in Van Putten \& Regimbau (2003) who
also find that the GRB peak luminosities and the beaming factors
are correlated.

We find a correlation coefficient between $L$ and $1/f_b$ of $\sim 0.5$.
The probability to obtain such a large value by chance,
for uncorrelated $L$ and $1/f_b$, is smaller than 0.05. 
The regression line slope
of log(L) vs log(1/f$_b$) is $\sim 0.8$.

The correlation is supported by a recent discovery of a relation 
between the observed peak energy, $E_p$, in the spectrum and L 
(Yonetoku et al. 2003).
This together with the correlation between $E_p$ and 
the isotropic equivalent energy in $\gamma$-rays, $E_{\rm iso}$, 
(Amati et al. 2002 and Lamb, Donaghy \& Graziani 2003) 
and with the Energy - angle relation
(Frail et al., 2001; Panaitescu \& Kumar 2001) imply $L_j=f_b L$.
However, within the
present sample there seem to be some spread in the value of $L_j$
as can be seen from Figure 4. To check the validity of our
analysis we take this spread into account. 

We first consider the case in which the distribution of $L_j$ 
is represented by a narrow-delta function and then discuss the effect 
of including the observed spread in $L_j$

\begin{figure}
%\centering \noindent
{\par\centering
\resizebox*{0.95\columnwidth}{!}{\includegraphics
%[width=8cm,height=8cm]
{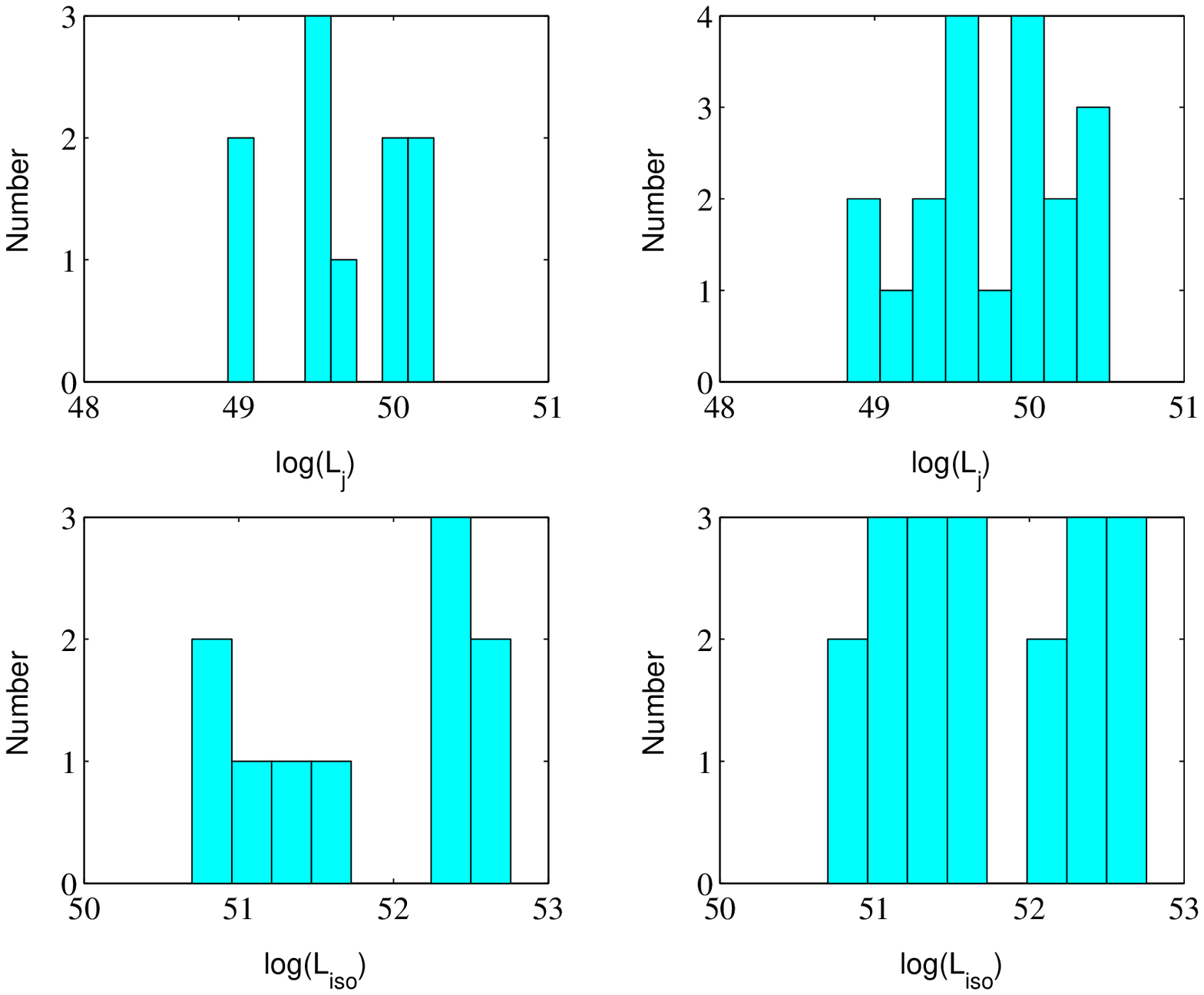}} \par}
\caption{\label{fig3} Left: the
distribution of the isotropic peak luminosity (lower panel) and
the distribution of the jet peak luminosity (upper panel)
of 10 GRBs in the BATSE catalog with angle determination. Right: the
distribution of the isotropic peak luminosity (lower panel) and
the distribution of the jet peak luminosity (upper panel) of
 a large sample of 19 GRBs with angle determination.}
\end{figure}

Now if the jet angle and the peak luminosity are related we can
derive the real distribution of the GRB jet opening angles. Let
$P(\theta,L)$ be the distribution of number of bursts with opening
angle $\theta$ and isotropic luminosity $L$. The observed
distribution of number of bursts with opening angle $\theta$ in
the small angles approximation can be written as:
%\begin{equation}
%\label{ptheta} P_{obs}(\theta) \sim  \int \frac{\theta^2}{2}
%P(\theta,L) dL \int_0^{z_{max}(\theta,P_{lim})}
%\frac{R_{GRB}(z)}{1+z} \frac{dV(z)}{dz}dz
%\end{equation}
\begin{eqnarray}
\nonumber
 P_{obs}(\theta) &=& \int f_b\,
P(\theta,L) dL\sim  \int \frac{\theta^2}{2}
P(\theta,L) dL \qquad\\
&&  \int_0^{z_{max}(\theta,P_{lim})} \frac{R_{GRB}(z)}{1+z}
\frac{dV(z)}{dz}dz \label{ptheta}
\end{eqnarray}
Given the correlation between the GRB peak luminosity and the
beaming factors we can assume that the condition $L(\theta^2/2)
=L_j$, with $L_j$ roughly constant for all the GRBs, holds. Hence
we have that $P(\theta,L)=\bar{P}(L)\delta(\theta-\sqrt{2
L_j/L})$. Using the fact that $\bar{P}(L)=L/L_j\Phi(L)$ we obtain
the intrinsic (corrected for beaming) angle distribution:
%Tsvi I have corrected the following Eq.
\begin{equation}
\label{ptheta4} P(\theta)= \tilde c_o \left\{
\begin{array}{ll}
\theta^{*2\beta} \theta^{-2\beta-1} & \theta < \theta^* \\
\theta^{*2\alpha} \theta^{-2\alpha-1} & \theta > \theta^* ,
\end{array}
\right.
\end{equation}
where  $\tilde c_o$ is a normalization constant so that the
integral over the angular distribution equals unity and $\theta^*
\equiv \sqrt{2L_j/L^*}$ is the break angle of the $\theta$
distribution.  $\theta^* \approx 0.12$rad with our canonical
parameters. This equation should be compared with $P(\theta)
\propto \sin\theta\sim\theta$ that is predicted by the  USJ model.
It is clear that the two distributions are inconsistent for large
angles.  A partial agreement is possible only for small angles
with $\beta \approx -2$. The disagreement at large angles is just
another indication of the fact that luminosity function implied by
the USJ jet model is incompatible with the observed peak flux
distribution at the low flux end (see Figure 1).

Substituting this relation in Eq.(\ref{ptheta}) and after some
algebra we find the observed $\theta$ distribution:
%\begin{equation}
%\label{ptheta4} P_{obs}(\theta)={ 2  c_o L_j \Phi(2 L_j/\theta^2)
%\over \theta^3}
% \int_0^{z_{max}(\theta,P_{lim})}
%\frac{R_{GRB}(z)}{1+z} \frac{dV(z)}{dz}dz
%\end{equation}
\begin{eqnarray}
\nonumber
 \label{ptheta5} P_{obs}(\theta)&=&{ 4  c_o L_j \Phi(2 L_j/\theta^2)
\over \theta^3} \qquad \qquad \qquad \qquad \qquad \\
&& \int_0^{z_{max}(\theta,P_{lim})} \frac{R_{GRB}(z)}{1+z}
\frac{dV(z)}{dz}dz.
\end{eqnarray}

\begin{figure}
%\centering \noindent
{\par\centering \resizebox*{0.95\columnwidth}{!}{\includegraphics
%[width=8cm,height=8cm]
{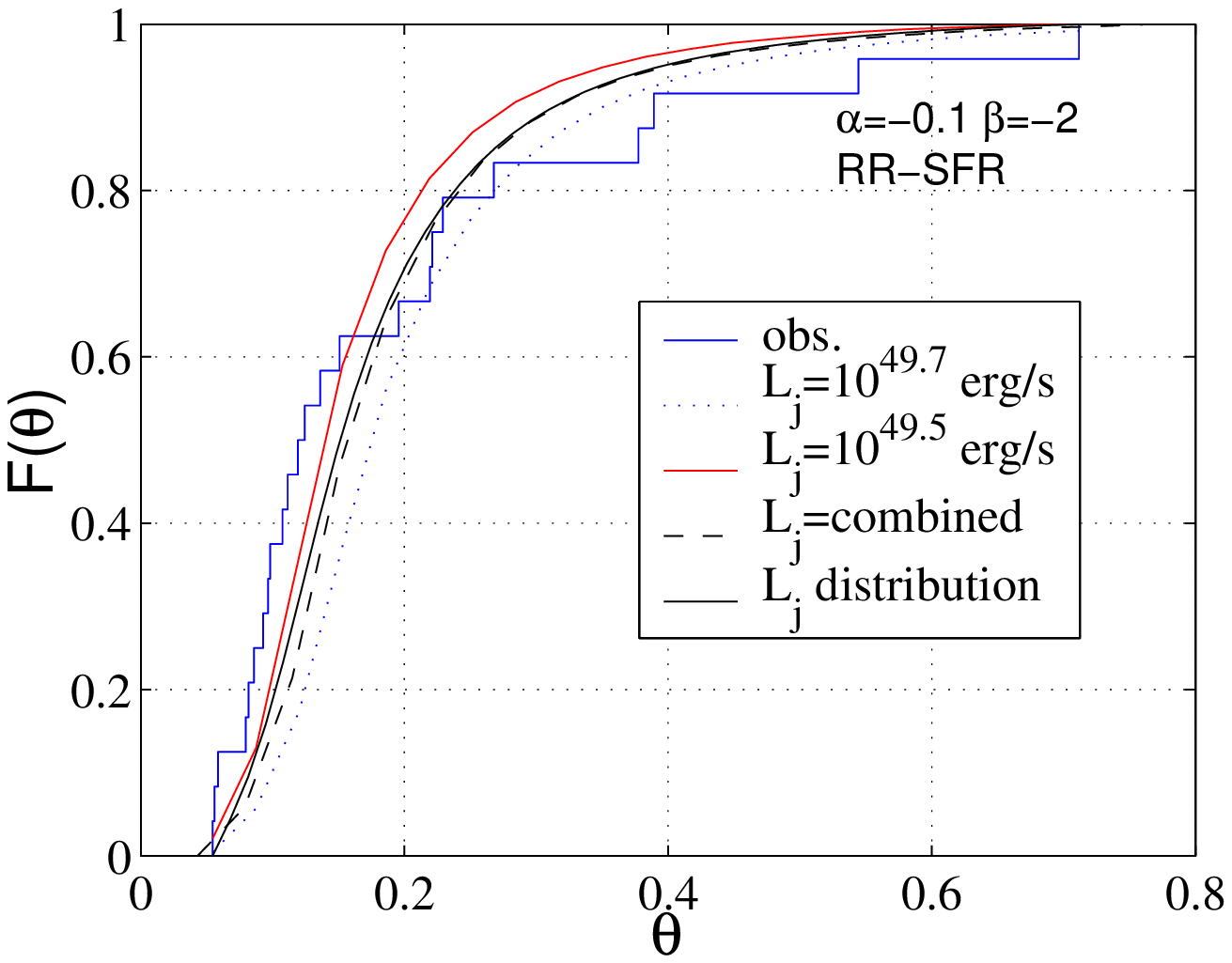}} \par}\caption{\label{fig4} The predicted
cumulative distributions of GRB opening angles for two
different values of $L_j$ and for a combination of the two
(reflecting the spread in the luminosity-angle relation) for
RR-SFR. These results are compared to the curve obtained
considering the observed distribution of $L_j$.
The histograms show the observed cumulative opening angle
distribution. }
\end{figure}

We compare the theoretical distribution with  a sample of 24 GRBs
with measured $\theta$ values (Blooom et al., 2003). We stress,
again, that selection effects are hard to quantify especially for
large and small opening angles. In Figure 5 we show the predicted
cumulative $\theta$ distribution derived from our luminosity
function with a RR-SFR and the observed one. We have chosen two
values of $L_j$ as representative of the distributions given in
Figure 4 (upper panel). We see that with these values of $L_j$ the
opening angle distribution reproduces quite well the observed one.

We have repeated the analysis with a distribution of
$\log_{10}(L_j)$ with finite width, $\sim 0.4$,
reflecting the spread in $L_j$ values seen in current data
(Figure 4).
The resulting $\theta$ distribution is
similar to the one that arises from  the combination of the two
values of $L_j$ (black solid line in Figure 5).

Once we know the angular distribution we can turn to our final
goal and estimate  the true GRB rate. To do so we multiply the
local observed rate, $\rho_0$ obtained in \S 2 by the correction
factor that now can be estimated directly from the luminosity
function given the relation between $f_b$ and $L$:
\begin{equation}
\label{eq:f_av} \langle f_b^{-1} \rangle=\frac{\int f_b^{-1}
\Phi_o(L)d\log L}{\int\Phi_o(L)d\log L} \approx \frac{\int L/L_j
\Phi_o(L)d\log L}{\int\Phi_o(L)d\log L}.
\end{equation}
$\langle f_b^{-1} \rangle$ is the beaming factor averaged over the
true, intrinsic, distribution of opening angles. We expect this
average to be smaller that the value derived by Frail et al.
(2001) due to the following reason: In the {\it intrinsic}
luminosity distribution   there are many bursts with a low
luminosity and large opening angles. These bursts dominate the
rate estimate and this should be factored in when calculating the
average rate correction (Sari, 2003). Put differently, in the {\it
observed} distribution, bursts with large $L$, i.e. large
$f_b^{-1}$, are over-represented compared to the true distribution
since they are observed to larger distances. In order to see this
explicitly, we note that the $\langle f_b^{-1} \rangle$ can be
derived directly from the observed distribution of GRB peak
luminosities without deriving $\Phi(L)$, using the following
argument. The total rate of GRBs (per unit $L$) in the observable
universe is $d\dot{N}/d\log L=\rho_0(L/L_j)\Phi(L) V_U$, where
$V_U$ is the effective volume of the observable universe, $V_U =
\int_0^{z_m} dz\,(dV/dz)R_{GRB}(z)/R_{GRB}(0)(1+z)$ ($z_m$ is the
maximum redshift from which signals arrive at us by today). The
observed rate is $d\dot{N}_{obs}/d\log L=\rho_0\Phi(L) V_L$, where
$V_L$ is the volume out to $z_L$, the maximum redshift out to
which bursts can be detected, $V_L = \int_0^{z(L)}
dz\,(dV/dz)R_{GRB}(z)/R_{GRB}(0)(1+z)$. The beaming factor,
defined in Eq.~\ref{eq:f_av}, can therefore be written as
\begin{equation}
\langle f_b^{-1} \rangle= \frac{\int
dL\,\frac{d\dot{N}_{obs}}{dL}\frac{L}{L_j} \frac{V_U}{V_L}}
{\int dL\,\frac{d\dot{N}_{obs}}{dL} \frac{V_U}{V_L}} .
\label{eq:f_av1}
\end{equation}
Eq.~\ref{eq:f_av1} explicitly reflects the fact that in
calculating the average beaming factor by which the observed local
rate $\rho_0$ should be multiplied in order to obtain the true
rate, $\langle f_b^{-1} \rangle$, the observed distribution of
bursts should be weighted by $V_U/V_L$. Since $V_U/V_L$ is
decreasing with $L$, a smaller weight is given to large $L$, i.e.
to large $f_b^{-1}$, than in the observed distribution.

Using Eq.~\ref{eq:f_av} with $L_j=10^{49.5}$~erg/s, we find
$\langle f_b^{-1} \rangle\simeq 53$ for SF2 and $\langle
f_b^{-1}\rangle\simeq 60$ for RR-SFR. 
These results do not change by inclusion of  the dispersion in
$L_j$.
As expected, these values
are lower than the value obtained by Frail et al. (2001), who
derived $\langle f_b^{-1} \rangle$ by averaging without the
$\propto V_L^{-1}$ weight. Using Eq.~\ref{eq:f_av1} with
$L_j=10^{49.5}$~erg/s and RR-SFR we obtain $\langle f_b^{-1}
\rangle\simeq 100$ for for the BATSE sample, and $\langle f_b^{-1}
\rangle\simeq 50$ for the larger sample. Using Eq.~\ref{eq:f_av1}
with $L_j$ determined for each burst from its estimated opening
angle, we find $\langle f_b^{-1} \rangle\simeq 70$ for for the
BATSE sample, and  $\langle f_b^{-1} \rangle\simeq 55$ for the
larger sample.  The values obtained using these different choices
are all within $\langle f_b^{-1} \rangle=75\pm25$. This agreement
reflects the fact that our derived $\Phi(L)$ gives a fair
representation of the data. The range of values obtained reflects
the range of uncertainty given current data.

We can also estimate the correction factor for the rate of GRB in
the case of the USJ model in the following way: The total flux of
GRBs per year (or per any other unit of time) is an observed
quantity that can be obtained by summing over the observed
distribution. In the USJ model the observed  energy-angle
relation implies: $E(\theta) = E_0/(\pi \theta^2)$ for
$\theta>\theta_c$ and $E(\theta)=E_0/(\pi \theta_c^2)$ for
$\theta<\theta_c$, where $\theta_{c}$ is the core angle of the
jet. The total energy that a burst with a USJ emits is:

%Tsvi I have corrected some typos here:

\begin{eqnarray}
E_{\rm USJ}& = &2  [\int_0^{\theta_c}(E_0/\theta_c^2)^2
\theta d\theta + \int_{\theta_{c}}^{\theta_{\rm max}} E_0
\theta^{-1}
d\theta  \nonumber \\
&=& E_0  [1+2 \log ( \theta_{\rm max}/\theta_c) ],
\end{eqnarray}
where $\theta_{max}$ is the maximal angle to which the jet
extends.

This immediately implies that
\begin{equation}
N_{\rm uniform} /N_{\rm USJ} =  [1+2 \log ( \theta_{\rm
max}/\theta_c) ]
\end{equation}
We don't know for sure what are the upper and lower limits but the
logarithmic dependance implies that the factor  cannot be smaller
than 2 or much larger than 5. This gives us the rate of uniform
jets to be about factor of 4 below the number of ``Uniform" jets.
Put differently this suggest that the correction factor for the
true rate of USJ compared to the rate of observed GRBs is a factor
of $\sim 20\pm 10$.

\section{Implications to Orphan Afterglows}

The realization that gamma-ray bursts are beamed with rather
narrow opening angles, while the following afterglow can be
observed over a wider angular range, led to the search for orphan
afterglows: afterglows that are not associated with observed
prompt GRB emission. The observations of orphan afterglows would
allow to estimate the opening angles and the true rate of GRBs
(Rhoads 1997).

Nakar, Piran \& Granot (2002) have estimated the total number of
optical orphan afterglows given a limiting magnitude using the
true rate of GRBs given by Frail et al. (2001). Nakar, Piran \&
Granot (2002) assumed, for simplicity, that all bursts have the
same opening angle, denoted $\theta^*$. In their canonical model
they assume that $\theta^*=0.1$rad while in the ``optimistic"
model they assume $\theta^*=0.05$rad. The smaller opening angle
gives of course more orphan afterglows. This analysis have to be
modified now using the new rate and the new correction factor that
we have found. It is clear that the assumption of a fixed opening
angle is not good enough and to obtain the rate of optical orphan
afterglows one have to perform another weighted average over the
observed $\theta$ distribution. This weight favors narrow jets
which produce orphan afterglows over a wide solid angle and for
which the rate correction is large. Overall (Nakar, 2003) one has
to correct downwards by a factor of 1.6 the rates of the
``canonical" model of Nakar, Piran \& Granot (2002). The orphan
afterglow rate obtained in the ``optimistic" model are clearly an
overestimate of the true rate, as this model assumed
$\theta^*=0.05$ which is significantly lower than the average
opening angle that we find here.

The situation is different for orphan radio afterglows, which are
seen at large angles. We re-estimate the number of orphan radio
afterglows associated to GRBs that should be detected in a
flux-limited radio survey (Levinson et al. 2002). Levinson et al.
have shown that the number of such radio afterglows detected over
all sky at any given time above a threshold  $f_{\nu,\rm min}$ at
1~GHz is
\begin{eqnarray}
\label{radio} \nonumber N_R\simeq 10\frac{\langle f_b^{-1}
\rangle}{70} \left(\frac{\rho_0}{0.5 \rm Gpc^{-3} yr^{-1}}\right)
\left(\frac{
f_{\nu,\rm min}}{5\rm mJy}\right)^{-3/2}
 \\
 \varepsilon_{e,-0.5}^{3/2}
 \varepsilon_{B,-1}^{9/8} n_{-1}^{19/24}
E_{51}^{11/6}.
\end{eqnarray}
Here, $E=10^{51}E_{51}$~ergs is the total fireball energy, assumed
equal for all bursts following Frail et al. (2001) and Panaitescu
\& Kumar (2001), $n=10^{-1}n_{-1}{\rm cm^{-3}}$ is the number
density of the ambient medium into which the blast wave expands,
and $\varepsilon_B=10^{-1}\varepsilon_{B,-1}$ ($\varepsilon_e=10^{-0.5}\varepsilon_{e,-0.5}$)
is the fraction of post shock thermal energy carried by magnetic field (electrons). For
$f_{\nu,\rm min}\sim5$~mJy, $N_R$ depends mainly on the local rate
$\rho_0$ and only weakly on the redshift evolution of the GRB
rate, since most of the detectable afterglows lie at low redshift
$z\lesssim 0.2$ (Levinson et al. 2002).

Afterglow observations imply a universal value of
$\varepsilon_e$ close to equipartition, $\varepsilon_{e,-0.5}\simeq1$, based on the clustering of explosion energies \cite{Frail01} and of X-ray afterglow luminosity
(Freedman \& Waxman 2001, Berger et al. 2003)
%\cite{Freedman01,Berger03a}.
The value of $\varepsilon_B$ is less well constrained by observations, since in most cases there is a degeneracy between $\varepsilon_B$ and $n$ in model predictions that can be tested by observations.
The peak flux of a GRB afterglow seen by an observer lying along the jet axis is proportional to $E_{\rm iso}(n\varepsilon_B)^{1/2}$, and for typical luminosity distance of $3\times10^{28}$~cm it is $\approx10(\varepsilon_B n/10^{-3}{\rm cm^{-3}})^{1/2}E_{\rm iso,54}$~mJy
(Waxman 1997, Gruzinov \& Waxman 1999, Wijers \& Galama 1999). Here, $E_{\rm iso}=10^{54}E_{\rm iso,54}$~erg is the isotropic equivalent energy. Observed afterglow fluxes generally imply $\varepsilon_B n\ge 10^{-3}{\rm cm}^{-3}$ for $E_{\rm iso,54}\sim0.1$, and values $\varepsilon_B n\sim 10^{-1}{\rm cm}^{-3}$ are obtained in several cases. We have therefore chosen the normalization $n_{-1}=\varepsilon_{B,-1}=1$ in Eq.~(\ref{radio}).

Using our value for $\langle f_b^{-1} \rangle$, $\langle f_b^{-1} \rangle=70$ instead
of $\langle f_b^{-1} \rangle=500$ given in Frail et al. (2001),
reduces the number of expected radio afterglows by a factor of
$\sim7$. However, the number of afterglows expected to be detected
by all sky $\sim1$~mJy radio surveys is still large, exceeding
several tens.

It should be pointed out that the lower limit on the beaming
factor inferred from the analysis of Levinson et al. (2002) is
unaffected by our modified value of $\langle f_b^{-1} \rangle$.
Assuming a fixed value of $f_b$ and expressing $N_R$ in terms of
the isotropic equivalent GRB energy, $E_{\rm iso.}=10^{54}E_{\rm
iso.,54}$~ergs, we have
\begin{eqnarray}
\nonumber
N_R \simeq 10^{4.5}f_b^{5/6} \left(\frac{\rho_0}{0.5 \rm
Gpc^{-3} yr^{-1}}\right) \varepsilon_{e,-0.5}^{3/2}\\
\varepsilon_{B,-1}^{9/8} n_{-1}^{19/24}
E_{\rm iso.,54}^{11/6}.
\label{eq:radio1}
\end{eqnarray}
Formally, the average value of $E_{\rm iso.}^{11/6}$ should appear
in this equation. We have chosen $<E_{\rm
iso.}^{11/6}>^{6/11}=10^{54}$~erg as a representative value. The
upper limit on the number of afterglows derived in Levinson et al.
(2002) implies a lower limit on the beaming factor, $f_b^{-1}>40$ for the parameter choice in Eq.~(\ref{eq:radio1}), indicating that radio surveys may
indeed put relevant constraints on the beaming factor.

\section{Conclusion}
In this work a combined L-$\theta$ distribution function of GRBs
is derived, $\Phi(L,\theta)$, for the uniform jet model. To this
aim we have rederived (following Schmidt, 1999) the luminosity
function,  $\Phi(L)$ by fitting the peak flux distribution and
used the relation $L\times \theta^2\sim$ const implied by the
observation on the sample considered by Bloom et al. (2003). We
have compared our results with those obtained in the framework of
USJ jet model (Perna et al. 2003) showing that  the luminosity
function implied by this model leads to a peak flux distribution
some what inconsistent with the observed one.

The luminosity function that best fits the observed peak flux
distribution is characterized by two power laws with slopes
$\alpha=-0.1$ and $\beta=-2$ and isotropic-equivalent break
luminosity $L^*\sim 7.1\times 10^{51}$ erg/s for a SF2 or $L^*\sim
4.4\times 10^{51}$ erg/s for a RR-SFR. Repeating the Schmidt
analysis we have found the observed local rate of long GRBs
$\rho_0 \sim 0.10$ Gpc$^{-3}$yr$^{-1}$ and $\rho_0 \sim 0.44$
Gpc$^{-3}$yr$^{-1}$ for the two SFR respectively. We have also
shown that with these luminosity functions we find a reasonable
agreement with the observed redshift distribution for a RR-SFR,
while a SF2 predicts too few bursts at low redshift. Nevertheless
since the redshift sample is strongly affected by selection
effects that are hard to estimate the possibility that GRBs follow
the SF2 cannot be ruled out by this analysis.

Using the sample of Bloom et al. (2003) we have shown that the
energy-angle relation (Frail et al., 2001; Panaitescu \& Kumar
2001)  applies also to the peak luminosity and that there is a
clustering of the true peak luminosity around $\sim 3.2\times
10^{49}$ erg/s. However the distribution is not a narrow
delta-function.  This implies that a luminosity function
determines a $\theta$ distribution as $L$ is related to $\theta$.
This is true even when the spread in  $L_j$ observed in the current
data is taken into account.
One important result of our work is that the resulting angular
distribution found under this assumption is consistent with the
observation.

We have re-calculated the correction to the rate of GRBs due to
beaming using a weighted average of the predicted angular
distribution instead of a simple average over the ``true'' angular
distribution as done by Frail et al. (2001) and by van Putten \&
Regimbau (2003). We find a correction factor $75\pm25$ (see the
end of \S~3). This is significantly smaller than the commonly used
correction factor, $\sim 500$, estimated by Frail et al. (2001) or
$\sim 475$ estimated by van Putten \& Regimbau (2003). . This
correction should also influence the estimated rates of both
optical (Nakar, Piran \& Granot, 2002) and radio (Levinson et al.,
2002) orphan afterglows (see \S~4).

The research was supported by the RTN ``GRBs - Enigma and a Tool"
and by an ISF grant for a Israeli Center for High Energy
Astrophysics. DG thanks the Weizmann institute for the pleasant
hospitality acknowledges the NSF grant AST-0307502 for financial support.
We thank Marteen Schmidt, Ehud Nakar and Cristiano Guidorzi for
helpful discussions.

\end{document}